\documentclass[draft,showpacs,preprintnumbers,amsmath,amssymb]{revtex4-1}
\newcommand{\fant}[1]{\phantom{#1}}
\newcommand{\be}{\begin{equation}}
\newcommand{\ee}{\end{equation}}
\newcommand{\wdg}{\wedge}
\newcommand{\ot}{\otimes}
\usepackage{multirow}

\begin{document}
\begin{abstract}
A system of field equations for an Einstein-Maxwell model with  $RF^2$-type nonminimal coupling
in a non-Riemannian spacetime with a nonvanishing torsion is derived and the resulting field equations are expressed in terms of  the Riemannian quantities based on  a metric with a Lorentzian signature. The torsion is generated by  the gradients of the electromagnetic field invariants. An electromagnetic constitutive tensor is introduced in the formulation of the  field equations.
\end{abstract}

\title{A nonminimally coupled Einstein-Maxwell model in a non-Riemann spacetime with torsion}


\author{Ahmet Baykal}
\email{abaykal@nigde.edu.tr}
\affiliation{Department of Physics, Faculty of Arts and Sciences, Ni\u gde University,  Bor Yolu,  51240 Ni\u gde, Turkey}

\author{Tekin Dereli}
\email{tdereli@ku.edu.tr}
\affiliation{Department of Physics, College of Science, Ko\c{c} University, Rumelifeneri Yolu, 34450 Sar\i yer, \.Istanbul, Turkey}
\date{\today}

\maketitle

\section{Introduction}

A spacetime with a nonvanishing torsion allows one to take the spin aspect of  particles into the account
in addition to the mass-energy aspect that is central to the  general  relativity theory (GRT) \cite{hehl-kerlick,shapiro}.
It is possible to  incorporate spinor matter fields into the geometry of spacetime by means of a propagating torsion \cite{hehl-mccrea-mielke-neemann}.
In this case, the torsion is regarded as the field strength of a gauge potential, that is, the coframe basis 1-forms whereas the
non-Riemannian connection is regarded as the gauge potential for a non-Riemannian  curvature. A non-Riemann spacetime geometry with a nonvanishing torsion is sometimes called a Riemann-Cartan geometry. Such theories constitute a  subclass of  matric-affine gravity
with  the linear affine connection in which both torsion and nonmetricity tensors play  essential roles. An  example of such a non-Riemannian model,
in the same spirit  as the current work,  was reported in  \cite{dereli-proca, vlachynsky-tresguerres-obukhov-hehl}  where the models involving
torsion and nonmetricity are related to a gauge potential for   a Proca field defined in a Riemannian spacetime with a Lorentz-signatured metric.
In such  extended gravitational theories with minimal coupling prescription, the torsion is related to  the spin  of matter fields,
as, for example, in the well-known case of the Dirac field with spin-$\frac{1}{2}$ in a non-Riemannian  spacetime with torsion.
In a Brans-Dicke type  scalar-tensor theory in which the scalar field couples  nonminimally to the gravity, it has been long known  \cite{dereli-tucker-PLB} that  the torsion can also be generated in terms of the gradient of the scalar field.

The modified gravitational models with a nonminimally coupled electromagnetic  field have been studied in quite different contexts with motivations  arising from
diverse subject matters. A generalized Einstein-Maxwell model with the nonminimal coupling of type $F\wdg F_{ab}*R^{ab}$ is related to the study of the
null electrovacuum solutions that are conformally flat \cite{prasanna} (the notation is introduced in the following section).
The general system of field equations for a large family of nonminimally coupled Einstein-Maxwell is presented for example in \cite{balakin-lemos}.
The complicated field equations in general involve higher derivatives in either metric components or in components of the Abelian gauge potential.
$RF^2$-type models with the field equations involving only the second order derivatives of the variables is constructed in \cite{horndeski}.

Although the nonminimal coupling of  the electromagnetic field is classically forbidden by the equivalence principle, they arise in the context of quantum electrodynamics in a curved spacetime background. For example, in the study of photon propagation in quantum electrodynamics in a curved background,
such nonminimal couplings arise in the effective Lagrangian \cite{drummond-hathrell,shore}. In the context of Kaluza-Klein
theories, the nonminimal couplings also arise from the dimensional reduction of Gauss-Bonnet and the general curvature-squared terms \cite{dereli-ucoluk,buchdahl,folkert} in higher dimensions.

More recently, $R^n F_{ab}F^{ab}$ type coupling has been studied to account for galactic magnetic fields \cite{campanelli,odintsov,kunze}.
In a more general setting, the Yang-Mills theory coupled to the $f(R)$ gravity has been studied in the
context of inflationary cosmology to account for the accelerated expansion of the Universe.
Exact solutions to the complicated field equations of various $RF^2$ type nonminimally coupled  models has been reported in \cite{folkert-sippel,balakin-bochkarev-lemos,gurses-halil,sert,dereli-sert-pp-wave}.

In the present work, a $RF^2$ type nonminimally coupled  Einstein-Maxwell  model  is studied in a non-Riemannian  spacetime with a nonvanishing toraion and in particular it is shown that the torsion is generated by  the gradients of the electromagnetic invariants $F\wdg *F$ and $F\wdg F$. It has previously  been reported  \cite{benn-dereli-tucker-gauge-invariance} that the gauge covariance related to an internal symmetry group is retained  in the context of non-Riemannian spacetime.  As will be apparent in the discussion of the $U(1)$-gauge covariance below, the geometrical description in terms of the exterior derivative and the algebra of differential  forms has certain advantages  over an equivalent formulation using  the non-Riemannian covariant derivatives and the tensorial components.

The layout of the paper is as follows. In the next section,
the geometrical preliminaries on non-Riemannian  spacetime with a nonvanishing torsion and vanishing non-metricity that will extensively be  used in the subsequent sections are introduced. In Sect. III, the field equations for the  nonminimally coupled model based on the Lagrangian density of the form $R(\alpha F\wdg *F+\beta F\wdg F)$ is derived using the variational techniques in terms of differential forms and the first order formalism with the coframe and connection 1-forms are treated as independent gravitational variables. Subsequently, the metric equations for non-Riemannian  spacetime are then rewritten in terms of the Lorentz-signatured Riemannian quantities. In Section IV,  derivation of  the field equations for the Riemannian case with a Lorentzian signature is briefly presented and the resulting equations are compared to those of the non-Riemannian  spacetime with a nonvanishing torsion derived in Sect. III. The paper ends with brief concluding comments.

\section{Geometrical preliminary}

In this section,  preliminary definitions of geometrical quantities and the notation for non-Riemannian spacetime with a  nonvanishing torsion, which  was studied long time ago by using the language of the algebra of differential forms \cite{trautman}, is presented in a self-contained way. The geometrical notation used here closely follows the one that is used in Ref. \cite{dereli-sert-pp-wave} with appropriate modifications.

The calculations in the following sections  will be carried out relative to a set   orthonormal basis coframe 1-forms $\{e^a\}$ for which the metric reads
$g=\eta_{ab}e^a\ot e^b$ with $\eta_{ab}=diag(-+++)$. The set of basis frame fields is $\{X_a\}$ and the abbreviation $i_{X_a}\equiv i_a$ is used for the contraction operator with respect to a frame field $X_a$. $*$ denotes the Hodge dual operator acting on basis 1-forms and
$*1=e^{0}\wdg e^1\wdg e^2\wdg e^3$ is the invariant volume element with a definite orientation. The abbreviations, for example,  of the form $e^{a}\wdg e^b\equiv e^{ab}$ for the exterior products of basis 1-forms are  extensively used for the convenience of the notation.
$F=\frac{1}{2}F_{ab}e^{ab}$ stands for the Faraday 2-form,
whereas $\tilde{F}\equiv *F=\frac{1}{2}\tilde{F}_{ab}e^{ab}$ is the dual Faraday 2-form.

The first structure equations  of Maurer-Cartan read
\be\label{cse1}
T^a
=
D(\Lambda)e^a
=
de^a+\Lambda^{a}_{\fant{a}b}\wdg e^b
\ee
where $D(\Lambda)$ stands for the covariant exterior derivative of the  connection non-Riemannian $\Lambda^{a}_{\fant{a}b}$. It is convenient to define contorsion 1-forms $K^{a}_{\fant{a}b}=K^{a}_{\fant{a}bc}e^c$ with $K_{ab}+K_{ba}=0$ in connection with the torsion 2-form. The total number of the independent components of the contorsion 1-form is  equal to that of torsion 2-form $T^a$ and contorsion 1-form  is related to the torsion 2-form by
\be\label{torsion-contorsion-def1}
T^a
=
K^{a}_{\fant{a}b}\wdg e^b.
\ee
This relation can be inverted to express the contorsion 1-form in terms of the contraction of the torsion 2-form as
\be\label{torsion-contorsion-def2}
K^{a}_{\fant{a}b}
=
-
\tfrac{1}{2}i^ai_b(T^{c}\wdg e_c)
+
i^a T_b
-
i_b T^a.
\ee
 The contorsion 1-form is especially useful in decomposing a non-Riemannian  connection into its Riemannian and post-Riemannian pieces in general. By combining the structure equation (\ref{cse1}) and the definition (\ref{torsion-contorsion-def1}), the non-Riemannian connection can be written  as the sum of a Riemannian part and a contorsion part as
\be\label{RC-connection-decomp}
\Lambda^{a}_{\fant{a}b}
=
\omega^{a}_{\fant{a}b}
+
 K^{a}_{\fant{a}b}
\ee
where the Levi-Civita connection $\omega_{ab}=-\omega_{ba}$ satisfies
\be\label{se1-zero-tor}
de^a+\omega^{a}_{\fant{a}b}\wdg e^b=0.
\ee
With respect to a vector field $X$, the covariant derivative  for the Levi-Civita connection $\omega^{a}_{\fant{a}b}$ can be defined by its action on the coframe basis 1-forms as $^\omega\nabla_Xe^a=-(i_X\omega^{a}_{\fant{a}b})e^b$. By making use of the definition of ${^\omega\nabla_X}$ in the structure equations (\ref{se1-zero-tor}), the relation between the operators $^\omega\nabla_{X_a}$  and exterior derivative  can be obtained as
\be\label{d-nabla-rel}
d=e^a\wdg {^\omega\nabla_{X_a}}
\ee
where both sides are operators acting on a $p-$form.
Moreover, for the connection $\Lambda^{a}_{\fant{a}b}$ Eq. (\ref{d-nabla-rel}) is modified to the form
\be\label{d-nabla-rel2}
d=e^a\wdg {^\Lambda\nabla_{X_a}}+T^a\wdg i_{a}
\ee
for the non-Riemannian covariant derivative $^\Lambda\nabla_{X_a}$ \cite{benn-tucker}.
In a non-Riemannian spacetime geometry with a nonvanishing torsion, the operator identity in Eq. (\ref{d-nabla-rel2})  is an essential ingredient  for the discussion \cite{benn-dereli-tucker-gauge-invariance} of the coupling of the  gauge fields with internal symmetries to gravity  in a gauge covariant manner. In the particular case of an Abelian gauge fields, the field strength is derived from a gauge potential 1-form $A=A_ae^a$ as $F=dA$ and it follows from the operator identity $d^2\equiv0$ that $F\mapsto F$ under $A\mapsto A+d\phi$ with $\phi$ being and arbitrary function. On the other hand, by making use of the operator relation (\ref{d-nabla-rel2}), it is possible to write the field strength in the alternate form as
\be\label{gauge-field-strength-exp2}
F=dA=e^a\wdg {^\Lambda\nabla_{X_a}}A+T^aA_a.
\ee
The well-known ``comma goes to semicolon" prescription  for the minimal coupling of a matter field involves only the covariant derivative  term in (\ref{gauge-field-strength-exp2}) that is not sufficient to ensure the $U(1)$-gauge covariance by itself \cite{straumann}.  In this regard, the second equality in Eq. (\ref{gauge-field-strength-exp2}) expresses how the minimal coupling prescription is to be altered in a non-Riemannian spacetime with a nonvanishing torsion when it is to be formulated in terms of a general non-Riemannian  covariant derivative. The spurious presence of the non-Riemannian  connection in (\ref{gauge-field-strength-exp2}) neither alters $U(1)$-gauge covariance nor imposes any constraint on the spacetime geometry.

As a consequence of the metric compatibility, a non-Riemannian  connection has the same antisymmetry property  as  the Levi-Civita  connection and the contorsion 1-forms: $\Lambda_{ab}=-\Lambda_{ba}$. Thus, the autoparallel curves defined by the non-Riemannian connection with a nonvanishing torsion coincide with the extremals of a Levi-Civita  connection only in the cases that torsion has some special properties \cite{hehl-kerlick}.

The second structure equations of Maurer-Cartan  defining a non-Riemannian  curvature 2-form  $R^{a}_{\fant{a}b}(\Lambda)$ corresponding to a non-Riemannian  connection $\Lambda^{a}_{\fant{a}b}$ with a nonvanishing torsion read
\be\label{RC-curvature}
R^{a}_{\fant{a}b}(\Lambda)
=
d\Lambda^{a}_{\fant{a}b}
+
\Lambda^{a}_{\fant{a}c}\wdg \Lambda^{c}_{\fant{a}b}
\ee
where  the curvature tensor components are defined as $R^{a}_{\fant{a}b}(\Lambda)\equiv\frac{1}{2}R^{a}_{\fant{a}bcd}(\Lambda) e^{cd}$ and, in general, it has
at most 36 independent components.
The decomposition of the non-Riemannian  connection (\ref{RC-connection-decomp})  then allows one to have an analogous decomposition for the curvature 2-forms $R^{a}_{\fant{a}b}(\Lambda)$ and its contraction.   Explicitly, by inserting the decomposition (\ref{RC-connection-decomp}) into the definition (\ref{RC-curvature}) one finds
\be\label{curvature-decomp1}
{R}^{a}_{\fant{a}b}(\Lambda)
=
R^{a}_{\fant{a}b}(\omega)
+
D(\omega) K^{a}_{\fant{a}b}
+
K^{a}_{\fant{a}c}\wdg K^{c}_{\fant{a}b}
\ee
where $R^{a}_{\fant{a}b}(\omega)$ is the curvature 2-form of the  Levi-Civita connection
\be
R^{a}_{\fant{a}b}(\omega)
=
d\omega^{a}_{\fant{a}b}
+
\omega^{a}_{\fant{a}c}\wdg \omega^{c}_{\fant{a}b}
\ee
and $D(\omega)$ denotes the covariant exterior derivative with respect to the Levi-Civita connection. The covariant exterior derivative with respect to the
non-Riemannian connection is denoted by ${D}(\Lambda)$ and it differs from $D(\omega)$ by an  appropriate contorsion term.
For example, the structure equation expressed in Eq. (\ref{curvature-decomp1}) can be rewritten in the form
\be\label{curvature-decomp2}
{R}^{a}_{\fant{a}b}(\Lambda)
=
R^{a}_{\fant{a}b}(\omega)
+
D(\Lambda) K^{a}_{\fant{a}b}
-
K^{a}_{\fant{a}c}\wdg K^{c}_{\fant{a}b}.
\ee

Because the non-Riemannian  connection $\Lambda^{a}_{\fant{a}b}$ is  metric compatible,
$D(\Lambda)\eta_{ab}=-\eta_{ac}\Lambda^{c}_{\fant{a}b}-\eta_{bc}\Lambda^{c}_{\fant{a}a}=0$, the integrability condition for the metric compatibility
retained as in the  Riemannian case with a Lorentzian signature: $D(\Lambda)^2\eta_{ab}=R_{ab}(\Lambda)+R_{ba}(\Lambda)=0$.

The first Bianchi identity, $D(\Lambda)^2 e^a=R^{a}_{\fant{a}b}\wdg e^b$, which are the integrability conditions for the first structure equations (\ref{cse1}), are modified and they explicitly read
\begin{align}
D(\Lambda)^2e^a
&=
R^{a}_{\fant{a}b}(\Lambda)\wdg e^b
\nonumber\\
&=
(D(\omega) K^{a}_{\fant{a}b}
+
K^{a}_{\fant{a}c}\wdg K^{c}_{\fant{a}b})\wdg e^b\label{reduced-1Bianchi-id}
\end{align}
and the result follows  with the help of the Bianchi identity $R^{a}_{\fant{a}b}(\omega)\wdg e^b=0$ and the curvature decomposition expressions (\ref{curvature-decomp1}) and (\ref{curvature-decomp2}). Note that the second line is also equal to $D(\Lambda)T^a$ by definition.
As a result,  the non-Riemannian curvature  tensor components $R^{a}_{\fant{a}b}(\Lambda)$ do not have the symmetry property of the Riemannian curvature tensor components $R_{abcd}(\omega)=R_{cdab}(\omega)$. Consequently, the Ricci tensor  components $\mathcal{R}_{ab}(\Lambda)$, which can be defined by the contraction of the curvature 2-form as $\mathcal{R}_{a}(\Lambda)\equiv i_b R^{b}_{\fant{a}a}(\Lambda)$ with $\mathcal{R}_{a}(\Lambda)\equiv\mathcal{R}_{ab}(\Lambda)e^b$,  are not necessarily  symmetrical. Moreover, the Einstein 3-form  corresponding to the non-Riemannian connection $\Lambda^{a}_{\fant{a}b}$ can be defined  in the same way  as it is defined in the Riemannian case with a Lorentzian signature as
\be\label{einstein-3form-def}
G^a(\Lambda)
\equiv
-\tfrac{1}{2}
R_{bc}(\Lambda)\wdg *e^{abc}
\ee
with $*^{-1}G_{a}(\Lambda)\equiv G_{ab}(\Lambda)e^b$ and  the components of Einstein tensor in terms of the components of the  Ricci tensor are $G_{ab}(\Lambda)\equiv \mathcal{R}_{ab}(\Lambda)-\frac{1}{2}\eta_{ab}{R}(\Lambda)$.

By using the expression (\ref{reduced-1Bianchi-id}) and the contraction of the first Bianchi identity,  it is possible to show that the antisymmetrical part of the Ricci tensor, $\mathcal{R}_{[ab]}(\Lambda)\equiv\frac{1}{2}(\mathcal{R}_{ab}(\Lambda)-\mathcal{R}_{ba}(\Lambda))$, has the expression
\be
\mathcal{R}_{[ab]}(\Lambda)
=
i_{a}i_bi_cD(\Lambda)T^c
\ee
in terms of the covariant exterior derivative of the torsion 2-form.

The second Bianchi identity  holds as in the Riemannian case with a Lorentzian signature for the metric: $D(\Lambda)R^{a}_{\fant{a}b}=0$.
The contracted second Bianchi identity expressed in terms of Einstein 3-form, on the other hand, takes the form
\be
D(\Lambda)G_a(\Lambda)
\equiv
-\tfrac{1}{2}\epsilon_{abcd} R^{bc}(\Lambda)\wdg T^d
\ee
where $\epsilon_{abcd}$ is the completely antisymmetric permutation symbol with $\epsilon_{0123}=+1$. Finally, note that the index raising and lowering commutes with the covariant exterior derivative of the non-Riemannian connection because the non-Riemannian connection  is assumed to be metric compatible.

\section{The Lagrangian and the field equations}

In four dimensions, the torsion 2-form can have at most 24 independent components, whereas the non-Riemannian curvature 2-form has at most 36  components.
Consequently, with regard to the number of independent components of the gravitational field variables in the non-Riemannian spacetime geometry with nonvanishing torsion, the total number of gravitational degrees of freedom is markedly larger  than the Riemannian spacetime with a Lorentzian signature. Consequently,  the increase in the number of the variables makes the study of the field equations in the non-Riemannian  spacetime difficult in general. However, a comparative study of a gravitational model  based on a non-Riemannian  spacetime geometry with a nonvanishing torsion may provide insight into the structure of the non-Riemannian  geometry in contrast to its  Riemannian counterpart with  a Lorentzian signature based on the same Lagrangian.

\subsection{The  auxiliary field definitions}

The discussion of the nonminimally coupled Einstein-Maxwell model will be based on the action integral $I$
\be
I
=
\int_ML[e^a, \Lambda^{a}_{\fant{a}b}, \lambda, F, \mu]
\ee
over a compact region $M\subset U_4$ on some chart on a Riemannian-Cartan manifold $U_4$.
The Lagrangian 4-form $L$ depends on the gravitational variables $\{e^a\}$ and $\{\Lambda^{a}_{\fant{a}b}\}$, the  Faraday 2-form $F$ and the Lagrange multipliers 0-form $\mu$. The Lagrangian 4-form $L=L_{E-M}+L_{nm}$ to be considered explicitly  reads
\be\label{lag-def}
L
=
\frac{1}{2\kappa}{R}_{ab}(\Lambda)\wdg *e^{ab}
-
\frac{1}{2}F\wdg *F
+
\lambda *1
+
\frac{1}{2}R(\Lambda)(\alpha F\wdg *F+\beta F\wdg F)
+
\mu\wdg  dF
\ee
where $\kappa\equiv 8\pi G $ is the Newton's gravitational constant with $c=1$, and
the Einstein-Maxwell Lagrangian $L_{E-M}$ with a cosmological constant $\lambda$  is extended by the fourth term on the right-hand side with the electromagnetic   invariants $F\wdg *F$ and $F\wdg F$ coupled to scalar curvature with the  coupling  constants $\alpha$ and $\beta$. ${R}(\Lambda)$ is the scalar curvature of the curvature 2-form ${R}^{a}_{\fant{a}b}(\Lambda)$ defined by $R(\Lambda)=i_a i_b{R}^{ba}(\Lambda)$. The last term on the right-hand side
is a Lagrange multiplier 1-form term  $\mu$  enforcing the constraint $dF=0$. This constraint ensures that 2-form $F$ can be derived from a local gauge potential 1-form $A$. Nonminimally coupled  Einstein-Maxwell models based on a more general  Lagrangian density than the one introduced in Eq. (\ref{lag-def}) have been studied recently in  the Riemannian context in Ref. \cite{sert}  also discussing the extension of such couplings to the case of non-Riemannian  spacetime with
a nonvanishing torsion.

It is convenient to rewrite  the Lagrangian density and the field equations that follow with reference to the Einstein-Maxwell model and the Maxwell's equations with the help of auxiliary tensors and $p$-forms. A particularly useful auxiliary tensor is  the constitutive tensor defined for the electromagnetic field.
In general all $RF^2$ couplings can be formulated compactly in terms of a suitable excitation 2-form, denoted by  $G$.
Explicitly, for a $RF^2$ coupling one can define the linear constitutive relation
\be
G=\mathcal{Z}(F)
\ee
with $\mathcal{Z}$  standing for a $(2,2)$-type constitutive tensor.
In this regard, the electromagnetic field equations, as well as the metric equations (\ref{RC-metric-eqns}), can be formulated in terms of the constitutive tensor defined by
\be\label{excitation-form-RC}
G
\equiv
F
-
R(\Lambda)(\alpha F-\beta \tilde{F}).
\ee
Accordingly, in terms of the 2-form $G$, the original Lagrangian 4-form in (\ref{lag-def}) can be rewritten in the form
\be\label{lag-def2}
L
=
\frac{1}{2\kappa}R*1+\lambda*1-\tfrac{1}{2}F\wdg *G+\mu\wdg dF.
\ee
It is also convenient for the notation to introduce the electromagnetic scalar invariants $X$ and $Y$  defined by the relations
\be\label{scalar-invariant-defs}
X\equiv -*(F\wdg *F),\qquad Y\equiv -*(F\wdg F),
\ee
respectively. It follows  from their  definition that the invariant scalars
can also be conveniently expressed as the contractions
\be
X=\tfrac{1}{2}F^{ab}F_{ab},\qquad Y=\tfrac{1}{2}F_{ab}\tilde{F}^{ab}
\ee
involving  the components of the Faraday 2-form and its dual. Because the definition of the excitation 2-form $G$ above involves the linear combination  of these scalar  invariants, it is also convenient to define a further scalar invariant:
\be
I\equiv\alpha X+\beta Y.
\ee

The expediency of the definitions of the constitutive tensor and, in particular, the excitation 2-form $G$ and the scalar invariants $X,Y,I$
in the formulation of the model will be evident in the following sections.

\subsection{The total variational derivative}

The Lagrangian studied here has no explicit connection 1-form dependence and $\Lambda^{a}_{\fant{a}b}$ enters into the  Lagrangian through the components of curvature tensor or else through the covariant derivative of spinor fields which are not considered in what follows.
Consequently, by using the variational identity $\delta {R}_{ab}(\Lambda)={D}(\Lambda )\delta \Lambda_{ab}$, and
the fact that the variational derivative $\delta$ commutes with exterior derivative $d$  one ends up with the general  result
\be\label{general-variational-id}
\delta L
=
\delta\Lambda_{ab}
\wdg
{D}(\Lambda)\frac{\partial {L}_{}}{\partial {R}_{ab}(\Lambda)}
+
\delta e^a
\wdg
\frac{\partial {L}_{}}{\partial e^a}
+
\delta F\wdg \frac{\partial L}{\partial F}
+
\delta \mu\wdg\frac{\partial L}{\partial \mu}
+
d\left(\delta\Lambda_{ab}
\wdg
\frac{\partial {L}_{}}{\partial {R}_{ab}(\Lambda)}\right).
\ee
The partial derivatives of the Lagrangian with respect to the forms  in the above expression can be expressed in terms
of partial derivatives of some appropriate functions with respect to tensor components \cite{kopczynsky}. Alternatively, the variational expression in Eq. (\ref{general-variational-id}) can be regarded as the formal definition of the derivative of the volume  form with respect to a $p$-form (See, for example, Ref. \cite{hehl-mccrea-mielke-neemann} for further details).

For the Lagrangian form given in Eq. (\ref{lag-def2}), it is possible to show that the total variational derivative  explicitly takes the form
\begin{eqnarray}\label{total-var-simplified}
\delta L
&=&
\delta e^a\wdg \left\{-\kappa^{-1}G_a(\Lambda)+\lambda*e_a+\tau_a[F,G]\right\}
+
\delta\Lambda_{ab}\wdg \tfrac{1}{2}\kappa^{-1}D(\Lambda)*e^{ab}
\nonumber\\
&&-
\delta F\wdg (d\mu+\tfrac{1}{2}*G)
+
\delta \mu\wdg dF
-
\tfrac{1}{2}\delta G\wdg *F
\end{eqnarray}
up to an omitted exact differential. The last term on the right-hand side of Eq. (\ref{total-var-simplified}) contributes to the electromagnetic field equations and   to the connection and coframe equations as well. The total variational derivative of this particular term is explicitly given by
\be
-
\tfrac{1}{2}\delta G\wdg *F
=
-
\tfrac{1}{2}\delta F\wdg *G
-
\delta e^a\wdg \tfrac{1}{2}I*\mathcal{R}_a(\Lambda)
+
\delta\Lambda_{ab}\wdg\tfrac{1}{2}D(\Lambda)I*e^{ab}.
\ee
The term $\tau_a[F, G]$ on the right-hand side in Eq. (\ref{total-var-simplified}) is the energy-momentum 3-form corresponding to the Lagrangian form. This term results from the variational derivative commuting with the Hodge dual operator in the term $-\frac{1}{2}F\wdg *G$ and it can conveniently  be expressed in  terms of   the Faraday and the excitation 2-forms as
\be\label{gen-asym-en-mom}
\tau_a[F, G]
\equiv
\tfrac{1}{4}(i_aF\wdg *G+i_aG\wdg *F-F\wdg i_a*G-G\wdg i_a*F)
\ee
analogous to the electromagnetic energy-momentum 3-form
\be\label{maxwell-en-mom-3form-def}
\tau_a[F]
=
\tfrac{1}{2}
(i_aF\wdg *F-F\wdg i_a*F)
\ee
corresponding to the Maxwell Lagrangian. By definition, one has the symmetry  relation $\tau_a[F, G]=\tau_a[G, F]$.
For the excitation 2-form defined in Eq. (\ref{excitation-form-RC}), the energy momentum 3-form (\ref{gen-asym-en-mom}) can also be reduced to the form
\be
\tau_a[F,G]
=
(1-\alpha R(\Lambda))\tau_a[F].
\ee
Eventually, one ends up with the expression
\begin{eqnarray}\label{total-der-final-form}
\delta L
&=&
\delta e^a\wdg \left\{-\kappa^{-1}G_a(\Lambda)+\lambda*e_a-\tfrac{1}{2}I*\mathcal{R}_a+\tau_a[F,G]\right\}
\nonumber\\
&&
+
\delta\Lambda_{ab}\wdg
\tfrac{1}{2}D(\Lambda)\left\{
(I+\kappa^{-1})*e^{ab}
\right\}
-
\delta F\wdg (d\mu+*G)
+
\delta \mu\wdg dF
\end{eqnarray}
for the total variational derivative of the Lagrangian. It is possible to read off  the explicit expressions for the partial derivatives in Eq. (\ref{general-variational-id}) from Eq. (\ref{total-der-final-form}) and thereby one can obtain  the field equations.

The complicated field equations for the nonminimally coupled model reduce to the  Einstein-Maxwell equations by setting $\alpha=\beta=\lambda=0$.
The model allows  the torsion to be generated by either of the scalar invariants $X, Y$, or more precisely, by the linear superposition of these invariants denoted by  $I$.

\subsection{The field equations for the coframe and the connection 1-forms}

Consequently, in terms of the excitation 2-form  $G$, the coframe (i.e. the metric) equations $\delta L/\delta e^{a}=0$ can be rewritten conveniently in the form
\be\label{RC-metric-eqns-const-form}
-
\kappa^{-1}{G}_a(\Lambda)
+
\lambda *e_a
-
I*\mathcal{R}_a(\Lambda)
+
\tau_a[F, G]=0.
\ee
The metric equations can also be written in a more explicit  form as
\be\label{RC-metric-eqns}
-
\kappa^{-1}{G}_a(\Lambda)
+
\lambda *e_a
-
(\alpha X+\beta Y) *\mathcal{R}_a(\Lambda)
+
(1-\alpha R(\Lambda))\tau_a[F]=0.
\ee
The electromagnetic  invariant $Y$ does not enter into the definition of the energy-momentum 3-form $\tau_{a}[F,G]$
since the expression $R(\Lambda)\beta F\wdg F $ in the definition of the scalar $Y$ does not involve the Hodge dual operator. Note, however,  that the term involving  the invariant $Y$ does contribute  to the  metric equations through a Ricci term.

Another feature of the energy-momentum form  $\tau_a[F, G]$ is that it is trace-free: $e^a\wdg \tau_a[F, G]=0$ for any excitation 2-form $G$
corresponding to most general $RF^2$ type couplings.
Consequently, the use of $\tau_a[F, G]$ naturally separates the trace part and the trace-free part  of the metric field equations for a given $RF^2$  model. The electromagnetic part contributing to the trace depends on a given $RF^2$ coupling, at hand. See,  for example, the expression given in Eq. (9) in \cite{dereli-sert-pp-wave} with the excitation 2-form $G=F-\gamma F_{ab}R^{ab}$.

The trace of the coframe equations (\ref{RC-metric-eqns}), that can be calculated by wedging (\ref{RC-metric-eqns}) by $e^a$, takes the form
\be\label{RC-trace}
R(\Lambda)\left(\kappa^{-1}*1-\alpha F\wdg *F-\beta F\wdg F\right)+4\lambda*1=0.
\ee

Next,  coming to the independent connection equations, $\delta L/\delta\Lambda_{ab}=0$ explicitly read
\be\label{RC-connection-eqn1}
{D}(\Lambda)[(I+\kappa^{-1})*e^{ab}]=0
\ee
and Eq. (\ref{RC-connection-eqn1}) provides an algebraic equation that can readily be solved for the torsion field in terms of the gradients of the scalar invariants   $X,Y$ by using   the identity
\be\label{hodge-torsion}
D(\Lambda)*e^{ab}
=
T^c\wdg *e^{ab}_{\fant{ab}c}.
\ee
Thus, it is convenient  to rewrite (\ref{RC-connection-eqn1}) with the help of  the identity given in Eq. (\ref{hodge-torsion}) and the definitions given in Eq. (\ref{scalar-invariant-defs}) as
\be\label{conn-sol}
dI\wdg *e^{ab}+(I+\kappa^{-1}) T^c\wdg *e^{ab}_{\fant{ab}c}=0.
\ee
In order to simplify this expression, it is convenient  to make use of the identity $e^a\wdg i_a\omega=p\omega$, that holds for an arbitrary $p$-form $\omega$. Consequently, one finds from Eq. (\ref{conn-sol}) that the torsion 2-form is given by
\be\label{torsion-expression}
T^a
=
-\tfrac{1}{2}(I+\kappa^{-1})^{-1} dI\wdg e^a
\ee
in terms of  the scalar invariant
$
I
=
\alpha X+\beta Y
$
defined above. The expression for the torsion 2-form (\ref{torsion-expression}) is analogous to the torsion generated by the Brans-Dicke  scalar field
in the Brans-Dicke theory \cite{dereli-tucker-PLB}. In contrast, the gradient of the Brans-Dicke scalar field is replaced with the gradient of the scalar invariant $I$ and, the torsion 2-form has  the trace part  with four components  as the only nonvanishing irreducible part \cite{hehl-mccrea-mielke-neemann}.
By making use of the relation (\ref{torsion-contorsion-def2}), the contorsion 1-forms corresponding to (\ref{torsion-expression}) can be found as
\be\label{contorsion-explicit-form}
K^{a}_{\fant{a}b}
=
-\tfrac{1}{2}(I+\kappa^{-1})^{-1}
\left(
i^adI e_b
-
i_bdI e^a
\right).
\ee

It follows from the torsion expression in Eq. (\ref{torsion-expression}) that the torsion is covariantly constant. Explicitly,  with the help of the expression for the contorsion 1-form, it is now straightforward to show that the exterior covariant derivative
\be
D(\Lambda)T^a
=
D(\omega)T^a+K^{a}_{\fant{a}b}\wdg T^b
\ee
vanishes identically with $T^a$ given in (\ref{torsion-expression}). Consequently, the Einstein tensor turns out to be symmetrical in the  particular model based on the Lagrangian 4-form (\ref{lag-def}).

\subsection{The electromagnetic  field equations}

The source-free    electromagnetic   equations that follow from $\delta L/\delta \mu=0 $ and $\delta L/\delta F=0 $ now can be written in terms  the excitation and the Faraday  2-forms conveniently as
\be
dF=0,\qquad d\mu+*G=0
\ee
respectively. With the help of the operator identity $d^2\equiv0$, the second equation can be written in the form $d*G=0$ which explicitly reads
\be\label{maxwell-eqn2}
[1+\alpha R(\Lambda)]d*F+ dR(\Lambda)\wdg  (\alpha*F+\beta F)=0.
\ee

Under certain topological assumptions, the field equation $dF=0$  allows one to define the field strength as $F=dA$ in terms of  a local gauge potential  1-form $A$ which is unique up to an exact 1-form by the Poincar\`{e} lemma \cite{straumann}.

To an  observer with a four velocity $U$ tangent to its world line inside the future light cone, the electric and magnetic induction 1-form fields associated with a given  Faraday 2-form can defined as
\be\label{e-b-def}
\mathbf{e}
\equiv
i_UF, \qquad \mathbf{b}=i_U\tilde{F}
\ee
respectively.
Conversely, by making use of $g(U,U)=-1$, Faraday 2-form can be rewritten in terms of $\mathbf{e}$ and $\mathbf{b}$ as
\be
F
=
\mathbf{e}\wdg \tilde{U}-*(\mathbf{b}\wdg \tilde{U}).
\ee
where $\tilde{U}$ stands for the  1-form associated with the vector field $U$ and relative to a coordinate basis one has the expansion
$\tilde{U}=g^{\mu\nu}U_\nu\partial_\mu$.
Similarly, electric displacement field $\mathbf{d}$ and  magnetic field $\mathbf{h}$ associated with the excitation  2-form $G$ are defined as
\be\label{d-h-def}
\mathbf{d}\equiv i_UG,\qquad \mathbf{h}\equiv i_U*G,
\ee
respectively. In terms of these 1-form fields defined relative to an observer, 2-form $G$ can similarly be rewritten in the form
\be
G
=
\mathbf{d}\wdg \tilde{U}-*(\mathbf{h}\wdg \tilde{U}).
\ee

In the general case, the definitions in Eqs. (\ref{e-b-def}) and (\ref{d-h-def}) then allow one to define the polarization and the magnetization 1-form fields as  $\mathbf{p}\equiv \mathbf{e}-\mathbf{d}$ and   $\mathbf{m}\equiv \mathbf{b}-\mathbf{h}$ respectively. For the excitation 2-form discussed above, the polarization and magnetization 1-forms  have the explicit forms
\begin{align}
\mathbf{p}
&=
R(\Lambda)i_U(\alpha F-\beta \tilde{F}),
\label{polarization}\\
\mathbf{m}
&=
R(\Lambda)i_U(\alpha \tilde{F}+\beta F),
\label{magnetization}
\end{align}
respectively, with a linear dependence on the non-Riemannian scalar curvature. The electromagnetic field equations formulated in terms of the excitation form   $G$ defined in Eq. (\ref{excitation-form-riem}) and the subsequent electromagnetic equations, for example,  the polarization and magnetization 1-forms can easily be decomposed into a Lorentz-signatured Riemannian part and a part induced  by the torsion field  with the help of excitation 2-form   $G$.

\subsection{The decomposition of the field equations}

The further scrutiny  of the $RF^2$ model in a non-Riemannian  spacetime with a nonvanishing torsion is motivated by the fact that the torsion can be eliminated from the coframe equation to rewrite it  in terms of the Riemannian quantities with a Lorentzian signature only and therefore identify the additional gravitational interactions induced by the torsional  degrees of freedom.

The expression (\ref{contorsion-explicit-form}) for the contorsion 1-form can be used to decompose the metric field  equations (\ref{RC-metric-eqns}) into the Riemannian quantities with a Lorentzian signature and the terms resulting from the torsion (\ref{torsion-expression}) with the help of the curvature identity (\ref{curvature-decomp1}). Explicitly, by making use of Eqs. (\ref{curvature-decomp1}) and (\ref{contorsion-explicit-form}), it is possible to obtain the decompositions
\begin{align}
G_a(\Lambda)
&=
G_a(\omega)
-
\frac{1}{(I+\kappa^{-1})} D(\omega)i_a*dI
+
\frac{3}{4(I+\kappa^{-1})^2}
\left\{
(i_a  dI)*dI+dI\wdg i_a  *dI
\right\},
\label{einstein-decomp-explicit}\\
R(\Lambda)*1
&=
R(\omega)*1
-
\frac{3}{(I+\kappa^{-1})}
d*d I
+
\frac{3}{2(I+\kappa^{-1})^2} dI\wdg *dI,\label{scalar-curvature-decomp-explicit}
\end{align}
for the Einstein 3-forms and the scalar curvature  in  the non-Riemannian spacetime, respectively.

The electromagnetic field equation $d*G=0$ also involves geometrical terms. In particular, the excitation 2-form can be written in terms of a term involving the scalar curvature $R(\omega)$ and the terms resulting from the torsion. Consequently, the electromagnetic field equations can easily  be decomposed  by using  Eq. (\ref{scalar-curvature-decomp-explicit}) in the definition (\ref{excitation-form-RC}). The decomposition of the scalar curvature $R(\Lambda)$ can  also
be used to decompose polarization and magnetization 1-forms  as well.

With the results (\ref{einstein-decomp-explicit}) and (\ref{scalar-curvature-decomp-explicit}) at hand, now it is possible to proceed in two ways:  One can insert these relations back into the original Lagrangian (\ref{lag-def}) to obtain  a reduced Lagrangian  density expressed in terms of Riemannian quantities with a Lorentzian signature and the interaction terms  induced by the torsion.  Equivalently, these relations can also be used to rewrite the corresponding field equations (\ref{RC-metric-eqns}). Following the second choice, it is possible to  find that the field equations (\ref{RC-metric-eqns})
can be rewritten in the form
\begin{align}\label{RC-reduced-metric-eqns}
&
-
\kappa^{-1}G_a(\omega)
+
\lambda*e_a
+
D(\omega)i_a*dI
-
I*\mathcal{R}_a(\omega)
+
(1- \alpha R(\omega))\tau_a[F]
\nonumber\\
&
+
\frac{3I}{2(I+\kappa^{-1})}\left(
i_ad*dI
-
\frac{i_a(dI\wdg *dI)}{2(I+\kappa^{-1})}
\right)
+
\frac{3\alpha}{I+\kappa^{-1}}
\left(
*d*dI
-
\frac{*(dI\wdg *dI)}{2(I+\kappa^{-1})}
\right)\tau_a[F]
\nonumber \\
&
-
\frac{3}{4(I+\kappa^{-1})^2}\{(i_adI)*dI+dI\wdg i_a*dI\}
=0
\end{align}
in terms of the Riemannian quantities with a Lorentzian signature. All the  terms in the second and the third lines in Eq. (\ref{RC-reduced-metric-eqns}) are induced by the torsion  2-form given in Eq. (\ref{torsion-expression}). The decomposition given in Eq. (\ref{RC-reduced-metric-eqns}) constitutes  one of the main results of the current work.

It has been remarked  in \cite{dereli-sert-pp-wave} that obtaining  a general explicit expression for torsion 2-form in terms of Faraday 2-form analogous to the one given in Eq. (\ref{torsion-expression}) by solving the connection equations is difficult for  the nonminimal couplings of the types $F\wdg F_{ab}*R^{ab}$, $F\wdg F_{ab}R^{ab}$, $F^a\wdg \mathcal{R}_{a}\wdg *F$ and $F^a\wdg \mathcal{R}_{a}\wdg F$ (or a general linear combination of these terms).
In this regard, the torsion in Eq. (\ref{torsion-expression}) provides an exceptional example among the $RF^2$ type couplings studied  in \cite{dereli-sert-pp-wave} previously.

Because the reduced coframe equation  (\ref{RC-reduced-metric-eqns}) is now expressed in terms of the Riemannian quantities with a Lorentzian signature plus the terms resulting from the algebraic torsion, it is possible to gain some insight into the mathematical structure of the model by  comparing it to the Einstein-Maxwell model that follows  from the Lagrangian density (\ref{lag-def}) in the Riemannian context with a Lorentzian signature. In this case, one introduces the vanishing torsion constraint on the independent connection in addition to the vanishing nonmetricity. The metric field equations in the Riemannian case with a Lorentzian signature that follow from the coframe variational derivative  of the Lagrangian (\ref{lag-def}) are obtained in the following section.

\section{A comparison to the Riemannian case with a Lorentzian signature}

For the above nonminimally coupled model, the field equations based on the Riemannian spacetime with a Lorentz signature can easily be recovered
by starting with  the coframe and connection 1-forms as independent gravitational variables and  then introducing  appropriate constraints into the first order formalism  \cite{sert,dereli-sert-pp-wave}. The vanishing torsion constraint on independent  connection 1-form, which is a dynamical constraint, can be introduced into the variational procedure consistently by  extending the original Lagrangian density  by the constraint term $\lambda_a\wdg T^a$. Explicitly, the Lagrangian (\ref{lag-def}) is to be extended by the constraint term of the form
\be
L_C[e^a, \Lambda^{a}_{\fant{a}b}, \lambda^a]
=
\lambda_a\wdg (de^a+\Lambda^{a}_{\fant{a}b}\wdg e^b)
\ee
where the Lagrange multipliers $\lambda^\alpha$ is now to be included in the  set of independent gravitational variables.
In this case, the extended Lagrangian $L_{ext.}=L+L_C$ has now the set of enlarged field variables: $\{e^a\}, \{\Lambda^{a}_{\fant{a}b}\}, \{\lambda^a\}, F, \mu$. It is a straightforward task to find that the total variational derivative of the constraint term with respect to its independent variables as
\be
\delta L_C
=
\delta e^a\wdg D(\Lambda)\lambda_a
-
\delta \Lambda_{ab}\wdg\tfrac{1}{2}(e^a\wdg \lambda^b-e^b\wdg \lambda^a)
+
\delta\lambda^a\wdg T_a
+
d(\delta e^a\wdg \lambda_a).
\ee
For the technical details of the total variational derivative  of the Lagrangian $L_{ext.}$ with respect to these variables  the reader is referred to \cite{sert} or \cite{dereli-sert-pp-wave}. It is possible to find that the connection equations $\delta L_{ext.}/\delta\Lambda_{ab}=0$ can be written in the form
\be\label{conn-eqn-Riem-case}
\Sigma^{ab}-\tfrac{1}{2}(e^a\wdg \lambda^b-e^b\wdg\lambda^a)=0
\ee
where the spin angular momentum 3-form  $\Sigma^{ab}$ can conveniently be expressed as
\be\label{Sigma-exp}
\Sigma^{ab}
\equiv
D(\Lambda)\frac{\partial L}{\partial R_{ab}}
=
\tfrac{1}{2}D(\Lambda)(I+\kappa^{-1})*e^{ab}.
\ee

In solving the connection equations (\ref{conn-eqn-Riem-case}) for the Lagrange multiplier 2-forms $\lambda^a$, the tensor-valued form
$\Sigma^{ab}$ expression is to be simplified  subject to the vanishing torsion constraint. In the notation used here, the constraint can be enforced simply  by
replacing $\Lambda_{ab}$ with $\omega_{ab}$ in the field equations.

The Lagrange multiplier $\lambda^a$ is a vector-valued 2-form and it has at most $24$ independent components in four dimensions. On the other hand, $\Sigma^{ab}=-\Sigma^{ba}$   is an antisymmetric (0,2)-tensor-valued 1-form and consequently has at most 24 independent components as well.
(Another  familiar example  of an analogous equivalence is the well-known equivalence between the torsion and the contorsion forms discussed in the preliminary section) Consequently, it is possible to express $\lambda^a$ in terms of $\Sigma^{ab}$ as
\be\label{general-lag-mul-exp}
\lambda^a
=
2i_b\Sigma^{ba}+\tfrac{1}{2}e^a\wdg i_bi_c\Sigma^{bc}
\ee
uniquely, by calculating two successive contractions of the connection equations (\ref{conn-eqn-Riem-case}).
Then, the general formula (\ref{general-lag-mul-exp}) can be used to derive the explicit expression
\be\label{LM-explicit}
\lambda^a
=
*(dI\wdg e^a)
\ee
 involving the gradients of the for electromagnetic invariants  $X,Y$ with $\Sigma^{ab}$  given as in (\ref{Sigma-exp}).
 After some manipulations involving contractions, this result readily follows from the expression given  in \cite{sert} for $\lambda^a$ by making appropriate  changes to take the invariant $Y$ into account. Subsequently,  by plugging the result (\ref{LM-explicit}) into the  metric field equations derived from coframe variations  $\delta L_{ext.}/\delta e^a=0$,  the metric equations then take the form
\be\label{metric-eqns-riemann-case}
-
\kappa^{-1}G_a(\omega)
+
\lambda *e_a
+
D(\omega)i_a*dI
-
I*\mathcal{R}_a(\omega)
+
\tau_a[F,G]=0
\ee
where the energy-momentum 3-form now has  the expression
\be
\tau_a[F,G]
=
(1-\alpha R(\omega))\tau_a[F].
\ee

By comparing the metric  field equations (\ref{metric-eqns-riemann-case}) with (\ref{RC-reduced-metric-eqns}) above,
one concludes  that the terms in the second and the third lines in (\ref{RC-reduced-metric-eqns}) result from the nonvanishing torsion
and   the terms in the first line in (\ref{RC-reduced-metric-eqns}), on the other hand, are expressed in terms of the Riemannian quantities with a Lorentzian signature only. In contrast to the analogous case of the nonminimally coupled scalar field, the effect of the contorsion forms is not simply a shift in the
multiplicative constant before the energy-momentum 3-form $\tau_a[F]$.

Finally, note that the electromagnetic  equations in  the Riemannian case with a Lorentzian signature for the metric  are formally the same as in the non-Riemannian  case  $dF=0$ and $d*G=0$ where now it is sufficient  to make  a  replacement in the definition of the excitation form $G$. In the
Lorentz-signatured Riemannian case, one can find that $G$ has the form
\be\label{excitation-form-riem}
G
=
F
-
R(\omega)(\alpha F-\beta  \tilde{F})
\ee
in terms of the scalar curvature corresponding to a Levi-Civita connection.
The expression for the excitation form given in Eq. (\ref{excitation-form-riem}) then can be used to  write down the electromagnetic equations. For example, the polarization and magnetization forms can be obtained by  replacing $R(\Lambda)$  with $R(\omega)$ in Eqs. (\ref{polarization}) and (\ref{magnetization}) respectively.

\section{Concluding comments}

The nonminimally coupled Einstein-Maxwell model studied above provides a new  example of
a non-Riemannian  geometry, in which the torsion is generated  in terms of  the gradients of the electromagnetic invariants retaining the  $U(1)$ gauge  invariance \cite{benn-dereli-tucker-gauge-invariance}. The modifications in the electromagnetic field equations are encoded into a suitable constitutive tensor linear in the curvature components and that such a constitutive tensor helps to rewrite the gravitational equations in a unified and compact form appropriate to general $RF^2$-type couplings. Furthermore, the metric field equations   of the nonminimally coupled model derived from the Lagrangian (\ref{lag-def}) can be decomposed into a Lorentz-signatured Riemannian part and a part resulting from nonvanishing torsion, which can be interpreted as gravitational interactions induced by torsion. Therefore, the current model provides an alternative to the previous works on the non-Riemannian  geometry  that relate torsion field to a minimally coupled spin-$\tfrac{1}{2}$ field. It is well known that a minimal coupling procedure  for spin-$\tfrac{1}{2}$ field generates an algebraic torsion  whereas a nonminimally-coupled spin-0 field also generates torsion \cite{dereli-tucker-PLB}. In this regard, the present work can be considered as  generalized non-Riemannian spacetime geometry  with a nonminimally-coupled spin-1 field generating the torsion.

It is, in principle, possible to carry out the analysis above for other $RF^2$ couplings considered in \cite{dereli-sert-pp-wave}, however it is  technically more difficult to obtain a closed expression for the corresponding torsion for  the  most general $RF^2$ couplings. The construction above provides some essential mathematical features of the general  $RF^2$ electromagnetic couplings in the context of  a non-Riemannian spacetime with a nonvanishing torsion.

The decomposition of the metric field equations into a Riemannian part plus a post-Riemannian part is a general feature in the geometry with a nonvanishing torsion and facilitates the  comparison between the extended gravity theories and the GRT in connection with their
mathematical structure as well as their physical interpretations and features embodied in the exact solutions these models.
In this regard, the field equations  (\ref{RC-reduced-metric-eqns}) may, for example, be helpful to  study the exact solutions of
complicated non-Riemannian gravitational models in relation to the analogous ones in  GRT.

Plane-fronted gravitational waves with parallel rays, so-called pp-waves,  provide an important example of ansatz that simplifies the field equations
(\ref{RC-reduced-metric-eqns}) to a manageable form. For the pp-wave metrical ansatz with a null electromagnetic field, both
invariants vanish and one has $I=0$.   By inspecting the metric equations (\ref{RC-reduced-metric-eqns})
for $I=0$, it is easy to see that they reduce to Einstein-Maxwell equations and in this case, the corresponding polarization and magnetization 1-forms defined vanish as well. Thus, one can readily deduce that the electrovacuum pp-waves are the solutions common in  GRT and the nonminimally coupled Einstein-Maxwell model discussed above.

Finally, note that it is possible to start the discussion above with  a slightly more general Lagrangian of the form $f(R)I*1$, or of the form $R f(I)*1$. However, such  Lagrangians only pose additional technical difficulties that can be carried out easily without offering further insight.
On the other hand, the particular type of $RF^2$ nonminimal coupling discussed above is singled out only by the relative technical simplicity among all the possible $RF^2$ couplings discussed in \cite{dereli-sert-pp-wave}. A more detailed investigation  of  the $RF^2$ couplings in full generality in the way as presented above is still an open problem.

\end{document}